\documentclass[11pt, letterpaper]{amsart}
\usepackage{textcomp}

\theoremstyle{definition}

\theoremstyle{remark}

\numberwithin{equation}{section}

\newcommand{\set}[1]{\left\{#1\right\}}

\newcommand{\prob}{\mathbb{P}}

\newcommand{\qprob}{\mathbb{Q}}

\newcommand{\dbra}[1]{[\kern-0.15em[ #1 ]\kern-0.15em]}
\newcommand{\dbraco}[1]{[\kern-0.15em[ #1 [\kern-0.15em[}
\newcommand{\dbraoc}[1]{]\kern-0.15em] #1 ]\kern-0.15em]}
\newcommand{\tC}{\widetilde{C}}

\newcommand{\Lb}{\mathbf{L}}

\begin{document}

\title[Free Lunch]{Free Lunch}%

\author{Constantinos Kardaras}%
\address{Constantinos Kardaras, Mathematics and Statistics Department, Boston University, 111 Cummington Street, Boston, MA 02215, USA.}%
\email{kardaras@bu.edu}%


\begin{abstract}
The concept of \emph{absence of opportunities for free lunches} is one of the pillars in the economic theory of financial markets. This natural assumption has proved very fruitful and has lead to great mathematical, as well as economical, insights in Quantitative Finance. Formulating rigorously the exact definition of absence of opportunities for riskless profit turned out to be a highly non-trivial fact that troubled mathematicians and economists for at least two decades. The purpose of this note is to give a quick (and, necessarily, incomplete) account of the recent work aimed at providing a simple and intuitive no-free-lunch assumption that would suffice in formulating a version of the celebrated Fundamental Theorem of Asset Pricing.
\end{abstract}

\maketitle


In the process of building realistic mathematical models of financial markets, absence of opportunities for riskless profit is considered to be a \emph{minimal} normative assumption in order for the market to be in equilibrium state. The reason is quite obvious. If opportunities for riskless profit were present in the market, every economic agent would try to reap them. Prices would then instantaneously move in response to an imbalance between supply and demand. This sudden price-movement would continue as long as opportunities for riskless profit are still present in the market. Therefore, in market equilibrium, no such opportunities should be possible.

The aforementioned simple and very natural idea has proved very fruitful and has lead to great mathematical, as well as economical, insight in the theory of Quantitative Finance. Formulating rigorously the exact definition of ``absence of opportunities for riskless profit'' turned out to be a highly non-trivial fact that troubled mathematicians and economists for at least two decades\footnote{The exact market viability definition is still sometimes the source of debate.}. As the road unfolded, the valuable input of the theory of stochastic analysis in financial theory was obvious; in the other direction, the development of the theory of stochastic processes benefited immensely from problems that emerged purely from these financial considerations.

\smallskip

Since the late seventies, it has been a folklore fact that there is a deep connection between absence of opportunities for riskless profit and the existence of a risk-neutral measure\footnote{Also called an Equivalent Martingale Measure.},
that is, a probability that is equivalent to the original one under which the discounted asset price processes has some kind of martingale property. Existence of such measures are of major practical importance, since they open the road to pricing illiquid assets or contingent claims in the market.
The above folklore result has been called the Fundamental Theorem of Asset Pricing.

The easiest and most classical way to formulate the notion of riskless profit is via the so-called \textsl{arbitrage strategy}
An arbitrage is a combination of positions in the traded assets that requires zero initial capital and results in nonnegative outcome with a strictly positive probability of the wealth being strictly positive at a fixed time-point in the future (after liquidation has taken place, of course). Naturally, the previous formulation of an arbitrage presupposes that a probabilistic model for the random movement of liquid asset prices has been set up. In \cite{MR540823}, a discrete-state-space, multi-period discrete time financial market was considered. For this model, the authors showed the equivalence between the economical ``No Arbitrage'' (NA) condition and the mathematical stipulation of existence of an equivalent probability that makes the discounted asset-price processes martingales.

Crucial in the proof of the result in \cite{MR540823} was the separating hyperplane theorem in finite-dimensional Euclidean spaces. One of the convex sets to be separated is the class of all terminal outcomes  resulting from trading and possible consumption starting from zero capital; the other is the positive orthant. The NA condition is basically the statement that the intersection of these two convex sets consists of only the zero vector.

\smallskip

After the publication of \cite{MR540823}, a saga of papers followed that were aimed, one way or another, at strengthening the conclusion by considering more complicated market models. It quickly became obvious that the previous NA condition is no longer sufficient to imply the existence of a risk-neutral measure; it is too weak. In infinite-dimensional spaces, separation of hyperplanes, made possible by means of the geometric version of the Hahn-Banach theorem, requires the closedness of the set $C$ of all terminal outcomes resulting from trading and possible consumption starting from zero capital. The simple NA condition does not imply this in general. This has lead Kreps in \cite{MR611252} to define a \textsl{free lunch} as a generalized, asymptotic form of an arbitrage.

Essentially, a free lunch is a possibly infinite-valued random variable $f$ with $\prob[f \geq 0] = 1$ and $\prob[f > 0] > 0$ that belongs to the \emph{closure} of $C$. Once an appropriate topology is defined on $\Lb^0$, the space of all random variables,  in order for the last closure (call it $\tC$) to make sense, the ``No Free Lunch'' (NFL) condition states that\footnote{$\Lb^0_+$ is the subset of $\Lb^0$ consisting of nonnegative random variables.} $\tC \cap \Lb^0_+ = \set{0}$. Kreps, in \cite{MR611252}, used this idea with a very weak topology on locally convex spaces and showed the existence of a \textsl{separating measure}\footnote{A separating measure is a probability $\qprob$ equivalent to the original one such that all elements of $C$ have nonpositive expectation with respect to $\qprob$. 
In the case of a continuous-time market model with locally bounded asset-prices, a separating measure automatically makes the discounted asset prices \emph{local} martingales --- this was proved in \cite{MR1304434}.}. However, apart from trivial cases, this topology does not stem from a metric, which means that closedness cannot be described in terms of convergence of sequences. This makes the definition of a free lunch quite nonintuitive.

\smallskip

After \cite{MR611252}, there were lots of attempts to introduce a condition closely related to NFL which would be more economically plausible, albeit still equivalent to NFL, and would prove equivalent to the existence of a risk-neutral measure. In general finite-horizon, discrete-time markets, it was shown in\footnote{For a compact and rather elementary proof of this result, see \cite{MR1837282}.} \cite{MR1041035} that the plain NA condition is equivalent to NFL. This seemed to suggest the possibility of a nice counterpart of the NFL condition for more complicated models.  Delbaen, in \cite{Del92}, treated the case of continuous-time, bounded and continuous asset prices and used a neat condition, equivalent to NFL, called\footnote{The appellation to this condition was actually coined by W. Schachemayer in \cite{MR1286705}.} ``No Free Lunch with Bounded Risk'' (NFLBR) that can be stated in terms of sequence convergence. Essentially, the NFLBR condition precludes asymptotic arbitrage at some fixed point in time, when the overall downside risk of all the wealth processes involved is bounded. Later, \cite{MR1286705} treated the case of infinite-horizon discrete-time models, where the NFLBR condition was
once again used. At this point, with the continuous-path and infinite-horizon discrete-time cases resolved, there seemed to be one more ``gluing'' step in order to reach a general version of the FTAP for semimartingale models. Not only did F. Delbaen and W. Schachemayer make this step for semimartingale models, they actually further weakened the NFLBR condition to the ``No Free Lunch with Vanishing Risk'' (NFLVR) condition, where the previous asymptotic arbitrage at some fixed point in time is precluded and the overall downside risk of all the wealth processes \emph{tends to zero} in the limit. In more precise mathematical terms, the NLFVR condition can be stated as $\overline{C} \cap \Lb^0 = \set{0}$, where $\overline{C}$ is the closure in the very strong $\Lb^\infty$-topology of (almost sure) \emph{uniform} convergence.

The NFLVR condition was finally the one that proved itself to be the most fruitful in obtaining a general version of the Fundamental Theorem of Asset Pricing; see \cite{MR1304434} and \cite{MR1671792}.
It is both economically plausible and mathematically convenient. Needless to say, and like many great results in science, the final simplicity and clarity of the result's statement came with the price that the corresponding proof was extremely technical.

\bibliographystyle{siam}
\bibliography{free_lunch}
\end{document}